# NiCl$_3$ Monolayer: Dirac Spin-Gapless Semiconductor and Chern Insulator

Junjie He,[†]  Xiao Li*,[§], Pengbo Lyu,[†] Petr Nachtigall*,[†]

[†]*Department of Physical and Macromolecular Chemistry, Faculty of Science, Charles University in Prague, 128 43 Prague 2, Czech Republic*

[§]*Department of Physics, University of Texas at Austin, Austin, TX, USA*

E-mail: lixiao150@gmail.com, petr.nachtigall@natur.cuni.cz

## ABSTRACT

The great obstacle for practical applications of the quantum anomalous Hall (QAH) effect is the lack of suitable QAH materials (Chern insulators) with large non-trivial band gap, room-temperature magnetic order and high carrier mobility. The Nickle chloride (NiCl$_3$) monolayer characteristics are investigated herein using first-principles calculations. It is reported that NiCl$_3$ monolayers constitute a new class of Dirac materials with Dirac spin-gapless semiconducting and high-temperature ferromagnetism (~400K). Taking into account the spin-orbit coupling, the NiCl$_3$ monolayer becomes an intrinsic insulator with a large non-trivial band gap of ~24 meV, corresponding to an operating temperature as high as ~280K at which the quantum anomalous Hall effect could be observed. The calculated large non-trivial gap, high Curie temperature and single-spin Dirac states reported herein for the NiCl$_3$ monolayer lead us to propose that this material give a great promise for potential realization of a near-room temperature QAH effect and potential applications in spintronics. Last but not least the calculated Fermi velocities of Dirac fermion of about $4 \times 10^5$ m/s indicate very high mobility in NiCl$_3$ monolayers.

## TOC

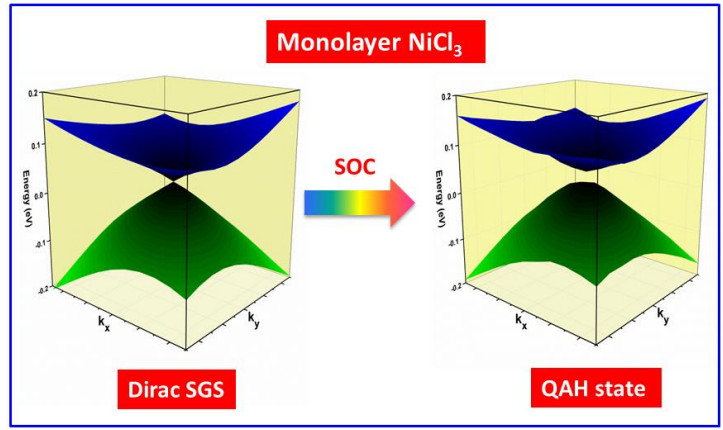

## 1. Introduction

Chern insulator or Quantum anomalous Hall insulator is a novel topological phase of matter characterized by a finite Chern number and the spin-polarized helical edge electron states within the bulk band gap.[1] Without an external magnetic field, internal magnetic exchange interaction (ferromagnetic and antiferromagnetic order) can breaks time-reversal symmetry (TRS) together with opening a non-trivial spin-orbital coupling induced gap, giving rise to a quantized anomalous Hall conductivity.[2] The helical edge states are robust against defects and impurities and thus such materials are attractive for highly promising applications in low power consumption electronic and spintronics devices.[2] Introducing magnetic order in topological insulators (TIs) to break the TRS such as chromium-doped $Bi_2Te_3$,[3] manganese doped HgTe quantum wells (QWs)[4] etc. are therefore expected to be promising route for realizing the QAH effect. Very recently, the QAH effect have been observed experimentally in Cr doped topological insulator $(Bi,Sb)_2Te_3$ film[5] in extremely low temperature (below 30 mK) due to a week magnetic coupling in doped Cr atoms and a small band gap. For practical applications it is crucial to search for QAH materials with the sizeable band gap, high Curie temperature ($T_c$), as well as the high carrier mobility.[6] Recently, a variety of QAH materials have been predicted by using impurities,[3,4] adatoms,[7] or chemical decorations[8] of graphene-based and Bi-based materials, and also Metal-organic-framworks,[9,10] interface structures[11,12] and heterostructure materials (*i.e.* $CrO_2/TiO_2$, $(Bi,Sb)_2Te_3/GdI_2$, and double perovskites)[13,14,15]. Most of these theoretically proposed materials for QHA effect show below the room-temperature $T_c$ due to the week magnetic order or a small SOC-gap.

In recent years, spin-gapless semiconductors (SGSs), exhibiting a band gap in one of the spin channels and a zero band gap in the other, have received considerable attention due to their unique electronic properties and potential applications in novel spintronic devices.[16] Dirac spin-gapless semiconductor (SGS) or Half semi-Dirac states, combining a single-spin massless Dirac fermions and half-semimetal with broken TRS, have been recently proposed for utilization of spin degree of electrons in Dirac materials.[17,18,19,20] By spin-orbit coupling (SOC), the gap opening may trigger QAH isolator transition in only one spin channel, which have been predicted to be in a few system, such as transition-metal intercalatition into epitaxial graphene on SiC(0001),[17] and $CrO_2/TiO_2$ herterstructure.[14] The search for new member for Dirac SGS for realization of QAH effect is of great importance for both a fundamental interests and practical applications.

Transition metal trichlorides (TMHs), a family of layered materials with the general formula ($TMCl_3$) have novel electronic and magnetic properties.[21,22,23,24,25] Among them, a relatively weakly interacting layers of a 3D $RuCl_3$ (dominated by van der Waals interactions) have been exfoliated into 2D materials from bulk phase recently [26] and the first-principle calculation demonstrates the $RuCl_3$ monolayer is metallic. In particular, Zhou et. al.[27] have recently indicated that the mixed metal chlorides ($NiRuCl_6$) are intrinsic half-metal antiferromagnets, which can lead towards the QAH effect in an antiferromagnetic order. Based on first principle calculations we found that the 2D $NiCl_3$ monolayer, as another member of TMH family, is intrinsic Dirac spin-gapless semiconductor with the high temperature ferromagnetism. When SOC is taken into account, a large gap opening is found to be 24 meV at the HSE06 level, giving rise to the quantum anomalous Hall states. We further confirm that the $NiCl_3$ monolayer has nontrivial topological Dirac-gap states characterized by a Chern number of C=-1 and chiral edge states. The physical origin of its QAH effect is due to both the intrinsic SOC and ferromagnetism of $NiCl_3$ monolayer. Our theoretical work leads us to proposal a pathway toward both the realization of a high temperature QAH effect and spintronics applications.

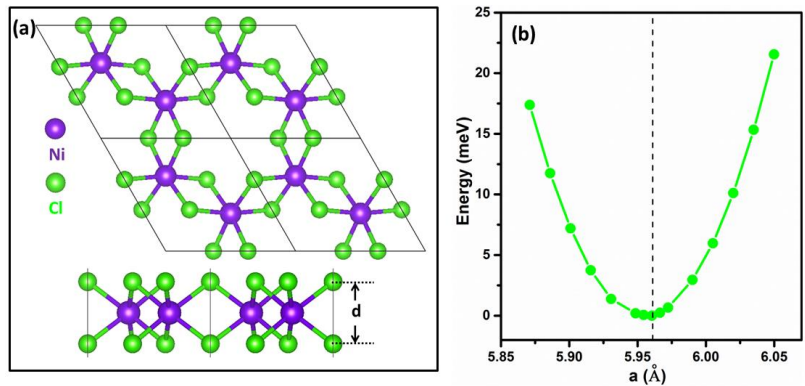

**Figure 1:** (a) The top and the side views of the optimized $NiCl_3$ monolayer. (b)Variation of total energy with the lattice constant.

## 3. Results and discussions

The structure of 2D $NiCl_3$ is shown in Figure 1a. Ni atoms form a 2d honeycomb lattice and each Ni atom bonded to six Cl atoms is in an octahedral environment. The dependence of total energy on the lattice constant of $NiCl_3$ monolayer is shown in Figure 1b. The lattice constant of 2D $NiCl_3$ is calculated to be 5.966 Å at the PBE level. The optimized bond length between Ni and Cl atoms $d_{Ni-Cl}$ is 2.60 Å. The vertical distance between two halide planes is calculated to be 2.93 Å. The 2D Young's modulus for NiCl$_3$ monolayer is calculated as:[17]

$$Y_{2D} = A_0 \left(\frac{\partial^2 E}{\partial A^2}\right)_{A_0} = \frac{1}{2\sqrt{3}}\left(\frac{\partial^2 E}{\partial a^2}\right)_{a_0}, \qquad (1)$$

where $E$ is the total energy per unit cell of NiCl$_3$, $a$ and $A$ stand for the lattice constant and surface area, respectively. Thus calculated 2D Young's modulus (Figure 1b) is estimated to be 25 N/m for NiCl$_3$ monolayer, which is very close to the value obtained previously for V- and Cr-based Chloride[23,24] and it is about 7% of the in-plane stiffness of the graphene (340 N/m). To further confirm the dynamic stability of NiCl$_3$ monolayers, its phonon spectra were calculated (Fig 2a). There is no negative frequency phonon in the whole Brillouin zone, indicating that the NiCl$_3$ monolayers are dynamically stable. Moreover, AIMD calculations carried out for 9 ps (with a time step of 3 fs) at 300K show clearly that the structure of NiCl$_3$ monolayer are nearly unchanged (Fig 2b), with the energy almost unchanged during the simulation, suggesting that the NiCl$_3$ monolayers are thermally stable at room temperature. Most importantly, the system remains magnetic throughout the simulation with an average magnetic moment of about 18 $\mu_B$ supercell (2 $\mu_B$ per unit cell) at 300 K, revealing that magnetic state of NiCl$_3$ is robust at the room temperature.

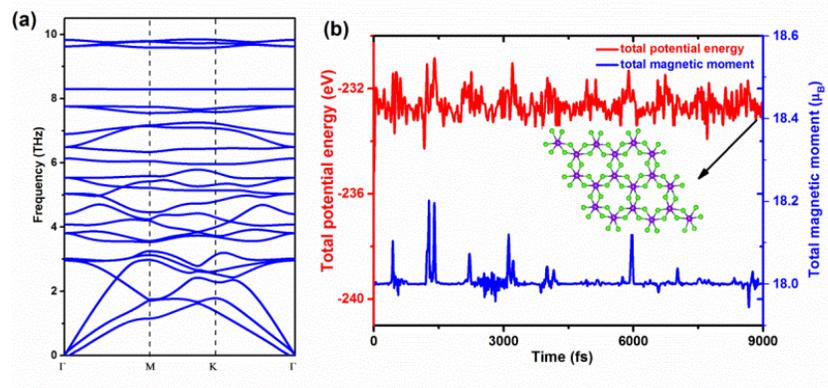

**Figure 2:** (a) Phonon band structures. (b) Total potential energy and total magnetic moment fluctuations of NiCl$_3$ as a function of simulation time (depicted in red and blue, respectively). The inset shows the corresponding structure after the simulation for 9 ps.

The NiCl$_3$ have spin-polarized ground states with a total magnetic moment of 2 $\mu_B$ per unit cell, corresponding to the d$^4$↑$^3$↓ spin configurations corresponding to Ni$^{3+}$, which can be verified by the Bader charge analysis.[28] To determine the preferred magnetic ground state structures of NiCl$_3$ systems, the collinear FM and AFM states are considered. The FM states are the most stable magnetic configurations. The nearest-neighbor exchange-coupling parameters $J$ (here the second and the third neighbor exchange-coupling are found to be one magnitude smaller than the nearest-neighbor) can be extracted by mapping the total energies of the systems with different magnetic structures to Ising model:

$$H_{spin} = -\sum_{i,j} J\, S_i \cdot S_j \quad , \qquad (2)$$

where $S$ is the net magnetic moment at the Ni site, $i$ and $j$ stand for the nearest Ni atoms. By mapping the DFT energies to the Heisenberg model, the $J$ can be calculated based on the energy difference between ferromagnetic and antiferromagnetic order by using expression: $J = \Delta E/6S^2$. Calculated exchange coupling parameters of NiCl$_3$ is 89.6 meV.

The Curie temperature $T_C$ is a key parameter for realization of the high temperature QAH effect and for spintronic applications. Based on the Weiss molecular-field theory (MFT), $T_C$ can be simply estimated by following formula:

$$T_C = \frac{2zJS(S+1)}{3k_B} \quad , \qquad (3)$$

where $z = 3$ is the number of nearest-neighboring Ni atoms in NiCl$_3$ monolayer, and $k_B$ is the Boltzmann constant. Following Eq. (3) a Curie temperature of 520 K for NiCl$_3$ monolayer has been obtained. Because of the possible overestimation of $T_C$ at the MFT level, MC simulations based on Ising model were also carried out. The MC simulation was performed on 80×80 2D honeycomb lattice using $10^8$ steps for each temperature. The specific heat capacity and magnetic moments are shown in Figure 3 as a function of temperature. It can be seen that the magnetic moment decrease to 0.8 $\mu_B$ at about 390 K and become 0 $\mu_B$ at 400 K. Therefore, the $T_C$ value for NiCl$_3$ monolayers is estimated to be about 400 K. Such temperature is orders of magnitude higher than current experimental temperature for the experimentally observed QAH effect. We propose that the NiCl3 monolayers can be a potential candidate for the high temperature QAH effect in spintronic applications.

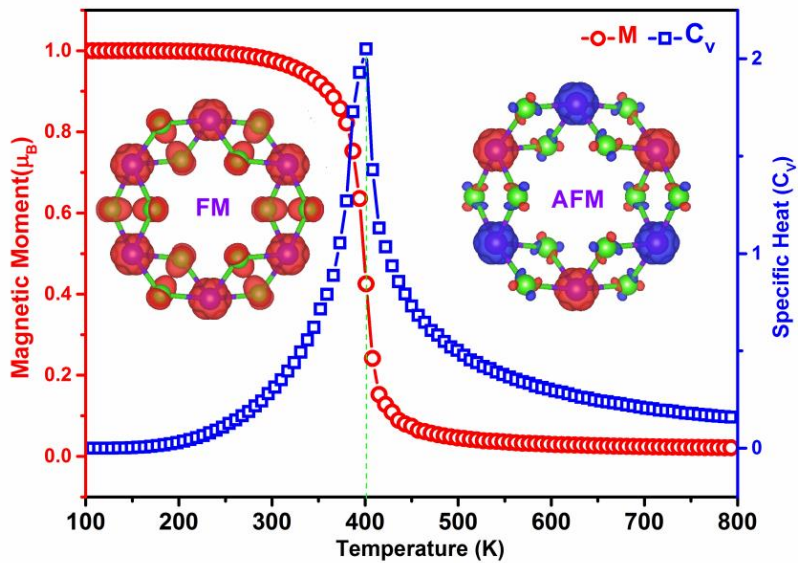

**Figure 3:** Variations of the average magnetic moment (red) and specific heat (blue) calculated for a NiCl$_3$ monolayer with respect to the temperature. Spin-polarized charge densities with spin directions for NiCl$_3$ are shown in insets (blue dotted areas depicts the spin up); iso-surface value of +/- 0.005 e/Å$^3$ shown.

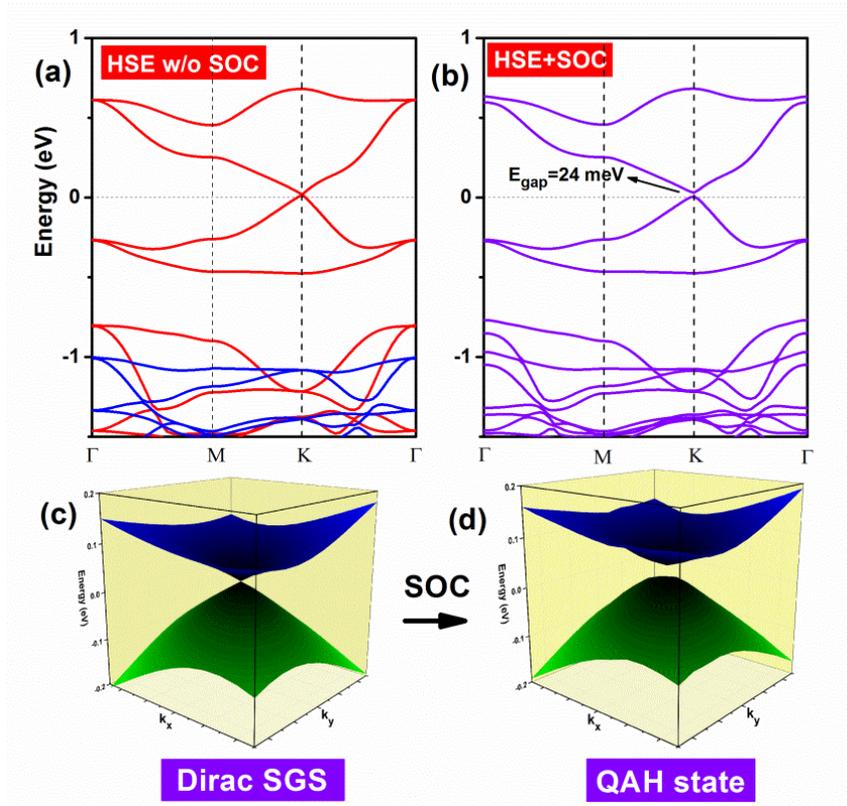

**Figure 4:** Band structures of 2D NiCl$_3$ (a) without SOC and (b) with SOC. The horizontal dotted lines indicate the Fermi level. The red and blue lines in (a) represent the spin up and spin down channels, respectively. The 3D band structures without(c) and with (d) SOC around the Fermi level in 2D *k* space are also presented.

The band structure of the NiCl$_3$ FM ground state is shown in Figure 4a. The spin-down channels of NiCl$_3$ possess a 4.09 eV band gap, whereas the spin-up shows a gapless semiconductor feature with a linear dispersion relation around the Fermi level. The spin-polarized massless Dirac fermions are found in the spin up channels of NiCl$_3$ at the high-symmetry *K* point of the Brillouin zone. The electronic structure of NiCl$_3$ show rather rare Dirac spin-gapless semiconductor, which can be used as a potential high-speed spin filter device. To further investigate the distribution of the linear dispersion relation in the Brillouin zone, the corresponding three dimensional band structure is also presented (Figure 4c).

The calculated Fermi velocities ($v_F$) of Dirac fermions are about $4 \times 10^5$ m/s for NiCl$_3$ monolayers, values that is approximately half of that found for graphene ($8 \times 10^5$ m/s). The existence of spin-polarized Dirac fermions in one spin channel and large band gap in another channel constitute a great potential for applications in spintronics.

Dirac materials, such as graphene, silicone, germanene, etc. are characterized by Dirac states composed of *p*-orbitals with week spin-orbital coupling. Thus, SOC opens just a tiny gap, making them Z2 topological insulator with TRS protected edged states. However, the Dirac states of NiCl$_3$ are mainly derived from the Ni-*d* orbital. The larger SOC gap of Ni-*d* orbitals with the broken TRS may lead to Chern insulator and to the QAH effect. The SOC gap was calculated by relativistic HSE06+SOC calculations to be 24 meV (Figure 4), which sufficiently large for the QAH effect at the temperature smaller than 280 K. The Chern insulater states of in NiCl$_3$ monolayer can be confirmed by the non-zero Chern numbers (C) calculated from the *k*-space integral of the Berry curvature ($\Omega(\bar{k})$) of all the states below the Fermi level using the formula of Kubo:[29,30,31]

$$C = \frac{1}{2\pi} \int_{BZ} \Omega(\bar{k}) d^2k \qquad (4)$$

$$\Omega(k) = \sum_{n<E_F} \sum_{m \neq n} 2 \,\text{Im}\, \frac{\langle \psi_{n\bar{k}} | v_x | \psi_{n\bar{k}} \rangle \langle \psi_{n\bar{k}} | v_y | \psi_{n\bar{k}} \rangle}{(\varepsilon_{m\bar{k}} - \varepsilon_{n\bar{k}})^2} \qquad . \qquad (5)$$

$\psi_{n\bar{k}}$ is the spinor Bloch wave function of band n with the corresponding eigenenergy $\varepsilon_{n\bar{k}}$. $v_x$ and $v_y$ are the i-th Cartesian components of velocity operator.

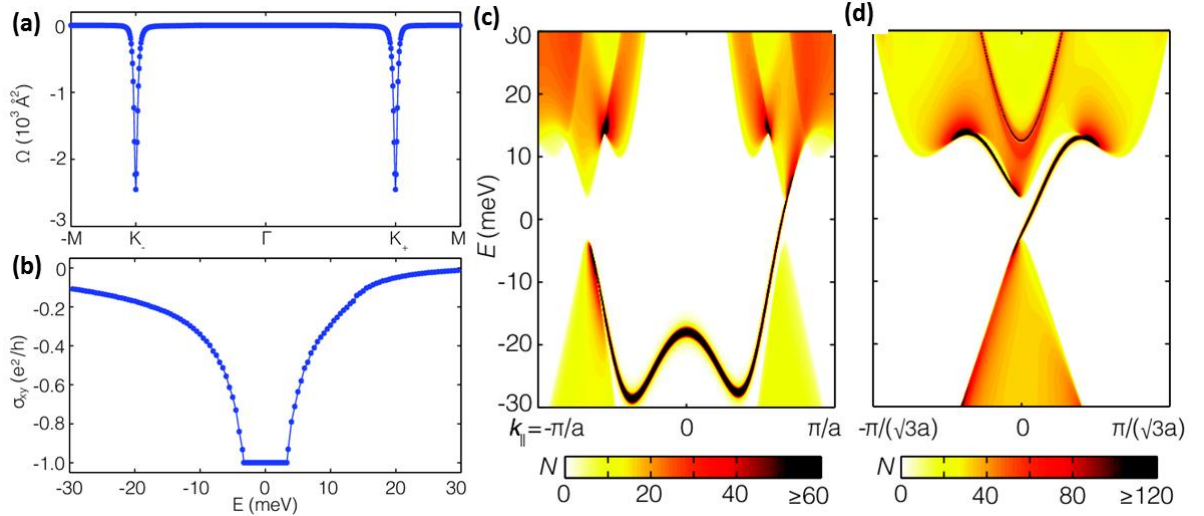

FIG. 5: (a) The distribution of the Berry curvature in momentum space for NiCl$_3$; (b) Anomalous Hall conductivity when the Fermi level is shifted around its original Fermi level; Calculated local density of states of edge states for (c) Zigzag and (d) armchair insulators. The edge states are calculated on the edge of a semi-infinite plane. The warmer colors (darker) represent higher local density of states at the edge.

The Berry curvature $\Omega(\bar{k})$ along the high-symmetry direction *M'-K-Γ-K'-M* shows two sharp spikes of the same sign located right at the *K* and *K'* points as shown in Fig 5a. By integrating the Berry curvature in the entire Brillouin zone, the calculated Chern number is *C* = -1 with a non-trivial topological states. As expected from the non-zero Chern number, the anomalous Hall conductivity shows a quantized charge Hall plateau of $\sigma_{xy} = Ce^2/h$ when the Fermi level is located in the insulating gap of the spin-up Dirac cone.

The existence of topologically protected chiral edge states is one of the most important consequences of the QAH state. To further reveal the nontrivial topological nature of NiCl$_3$ monolayer, we calculate the edge states of NiCl$_3$ monolayer with zigzag and armchair by Green's function based on Wannier functions obtained from PBE calculations, which reduces the cost of calculation and do not change the topology of electronic structure, besides a smaller band of 7 meV. As shown in Fig. 5, the nontrivial edge states (dark line) connecting the valence and conduction bands cross the insulating gap of the Dirac cone. The appearance of only one chiral edge state is consistent with the calculated Chern number $C = -1$, confirming the nontrivial topological nature of NiCl$_3$ monolayer. Using 24 meV equivalent to 280 K as a rough estimate, the QAH effect in NiCl$_3$ is expected to be robust below 280 K. This is much higher than the QAH experimental temperature (<100 mK) for a Cr doped Bi$_2$Se$_3$ film.[5] Furthermore, the FM ordering temperature as high as 400 K for NiCl$_3$ is large enough to retain the QAH phases in the above-mentioned temperature ranges. The single spin Dirac fermion mediated topological properties shows the NiCl$_3$ monolayer is the potential to generate the QAH effect.

Finally, the PDOS and orbital-projected band structure around Fermi level were calculated for NiCl$_3$ monolayer to gain insight into the origin of Chern insulator (Figure 6). The states near Fermi level have main contribution from the $d_{xy}$, $d_{x^2-y^2}$, $d_{xz}$ and $d_{yz}$ orbitals while the $d_{z^2}$ orbital does not contribute significantly. Therefore only $d_{xy}$, $d_{x^2-y^2}$, $d_{xz}$ and $d_{yz}$ orbitals are presented in the orbital-projected band structure comparing HSE and HSE+SOC levels. Without SOC, the states of fermi level are mainly contributed by the mixed $d_{xy}$, and $d_{xz}$ orbitals, while only a little weight of both $d_{yz}$ and $d_{x^2-y^2}$ orbitals contributes to the states near the Fermi level. When the SOC is considered, the gap opens but the states around Fermi level are dominated by the contribution mainly from $d_{yz}$ orbital. The $d_{yz}$ orbital between CB and VB bands is clearly separated each other as Dirac point as shown in Fig. 6b, opening the global energy gap. The separation plays a key role in the reversal of $d_{xz}$ and $d_{yz}$, leading to the Chern insulator in NiCl$_3$ monolayer. Such changing of orbital weight for $d$ states is similar with the previously reported antiferromagnetic Chern insulator NiRuCl$_6$ sheet.[27]

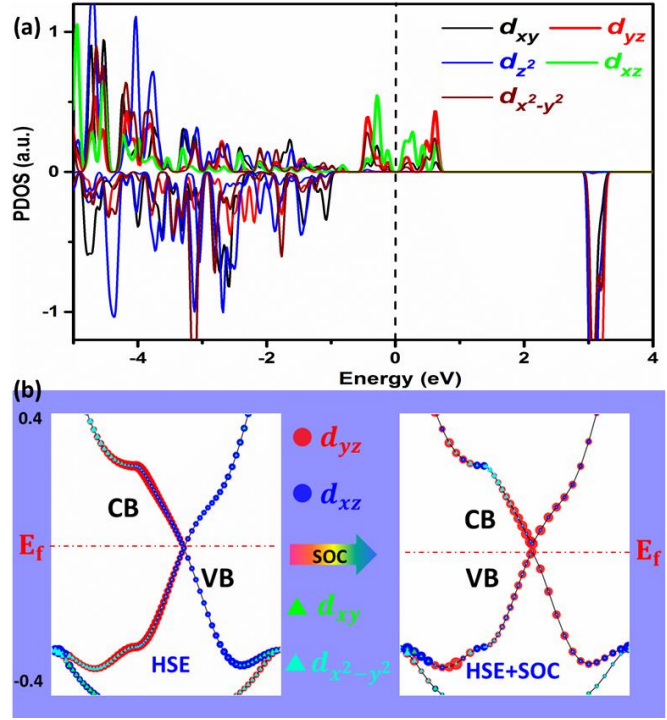

Figure 6: (a) The PDOS of Ni atoms calculated at HSE06 level. (b) The evolution of orbital-resolved band structures of NiCl3 monolayer without and with SOC.

## 4. Conclusions

The DFT calculations were used in a systematic investigation of the stability, and electronic and magnetic structures of $NiCl_3$ monolayers. The phonon calculations and ab initio molecular dynamics simulations suggest that single layers of $NiCl_3$ are dynamically ad thermally stable.. The $NiCl_3$ monolayers show the Dirac spin-gapless semiconducting characteristics and a high-temperature ferromagnetism. The Monte Carlo simulations based on the Ising model demonstrate that the Curie temperature of $NiCl_3$ monolayer is estimated to be as high as 400 K. In addition, a Fermi velocity ($v_F$) in $NiCl_3$ monolayer is calculated to be $4 \times 10^5$ m/s, which is comparable to graphene ($8 \times 10^5$ m/s). Taking spin-orbit coupling into account, the $NiCl_3$ monolayer becomes an intrinsic quantum anomalous Hall insulator with a large non-trivial band gap of about 24 meV, corresponding to an operating temperature of 280K. The large non-trivial gap, high Curie temperature and single-spin Dirac states for NiCl$_3$ monolayers give a rise to great expectation for both the realization of near room temperature QAH effect and potential application in spintronics.

## Methods and computational details

All calculations were performed using the Vienna *ab initio* simulation package (VASP)[32,33] within the generalized gradient approximation, using the Perdew-Burke–Ernzerhof (PBE) exchange-correlation functional.[34] Interactions between electrons and nuclei were described by the projector-augmented wave (PAW) method. The criteria of energy and atom force convergence were set to $10^{-6}$ eV and 0.001 eV/Å, respectively. A plane-wave kinetic energy cutoff of 500 eV was employed. The vacuum space of 15 Å along the NiCl$_3$ normal was adopted for calculations on monolayers. The Brillouin zone (BZ) was sampled using 15×15×1 Gamma-centered Monkhorst-Pack grids for the calculations of relaxation and electronic structures. All electronic structure of the 2D NiCl$_3$ is employed the hybrid HSE06 functional. Furthermore, to exam the thermal stability of the NiCl$_3$, the *ab initio* molecular dynamics (AIMD) simulations at 300 K in a canonical ensemble are performed using the Nosé heat bath approach. A 3×3×1 supercell of NiCl$_3$ monolayer was used in MD simulations. The phonon frequencies were calculated using a finite displacement approach as implemented in the PHONOPY code, in which a 2×2×1 supercell and a displacement of 0.01 Å from the equilibrium atomic positions are employed.[35,36] The electronic properties of the NiCl$_3$ monolayer obtained by VASP have been further reproduced by QUANTUM ESPRESSO package[37], with the norm-conserving pseudopotentials from the PS library[38] and a 120 Ry plane wave cutoff. Based on the Wannier functions obtained from the first-principles calculations in QUANTUM ESPRESSO,[39,40,41] we construct the edge

Green's function of the semi-infinite $NiCl_3$ monolayer. The edge spectral density of states, computed by the edge Green's function, shows the energy dispersion of edge states.[42] Berry curvature and the anomalous Hall conductivity are also calculated by Wannier interpolation.[36]

## Acknowledgments

This work was funded by the Czech Science Foundation Grant No. P106/12/G015 (Centre of Excellence). The calculations were partially performed at MetaCentrum and CERIT-SC computational facilities (MSM/LM 2010005 and OP VaVpI CZ. 1.05/3.2.00/08.0144). J.H. and P.L. acknowledge support from STARS Scholarship of Faculty of Science, Charles University in Prague.